\documentclass[11pt,english]{article}
\usepackage[T1]{fontenc}
\usepackage[latin9]{inputenc}
\usepackage{geometry}
\usepackage{amstext}
\usepackage{amssymb}
\usepackage{graphicx}
\usepackage{setspace}
\makeatletter
\newcommand{\lyxaddress}[1]{
\par {\raggedright #1
\vspace{1.4em}
\noindent\par}
}

\makeatother

\usepackage{babel}
\begin{document}

\title{Ring conformations in bidisperse blends of ring polymers}

\author{M. Lang$^{1}$}

\maketitle

\lyxaddress{$^{1}$Leibniz Institut für Polymerforschung Dresden, Hohe Straße
6, 01069 Dresden, Germany.}
\begin{abstract}
The size of rings (also called cyclic polymers) in bidisperse blends
of chemically identical rings is analyzed by computer simulations.
Data of entangled ring blends and blends of interpenetrating rings
are compared and it is shown that the compression of entangled rings
can be explained by the changes in the penetrable fraction of the
minimal surface bounded by the ring. Corrections for small rings can
be approximated by a concatenation probability $1-P_{OO}$ that a
ring entraps at least one other ring. Both results are in line with
a previous work \cite{Fischer} to explain the compression of entangled
rings in monodisperse melts. Bond-bond correlations in melts of interpenetrating
rings lead to similar corrections for ring sizes as reported previously
\cite{wittmer2007intramolecular} for monodisperse linear melts. For
entangled rings, bond-bond correlations show an anti-correlation peak
at a curvilinear distance of about ten segments that coincides with
a horizontal tangent in the normalized mean square internal distances
along the ring. Both observations become independent of melt molecular
weight for sufficiently large degrees of polymerization and such behaviour
is not found in samples with entanglements switched off. In consequence,
the length scale of topological interactions (entanglement length)
in a melt of entangled rings must be considered as constant in contrast
to a recent proposal by Sakaue \cite{Sakaue2012}. 
\end{abstract}

\section{Introduction}

Recently, it was suggested \cite{Fischer} that the conformations
of non-concatenated rings in monodisperse melt may follow four different
regimes as function of the degree of polymerization $N$. These regimes
are separated by three characteristic degrees of polymerization, $N_{OO}$,
$N_{C}$, and $N^{*}$. $N_{OO}$ describes the onset of the effect
of non-concatenation below which topological effects are not important,
$N_{C}$ is the cross-over between weak and strong compression of
rings, and $N^{*}$ is defined by the cross-over from a non-concatenation
contribution $f_{n}\sim\phi R^{2}$ to an overlap dominated concatenation
contribution $f_{n}\sim\phi N^{1/2}$ at $N>N^{*}$.

Below $N_{OO},$ the rings are uncompressed and ring conformations
nearly ideal. For $N_{OO}<N<N_{C}$, ring polymer conformations may
be described by a free energy of form

\begin{equation}
\frac{\Delta F}{kT}\approx b\phi R^{2}+\mbox{v}_{e}\frac{N^{2}}{R^{3}}.\label{eq:FET}
\end{equation}
Here, the first term with polymer volume fraction $\phi$ and root
mean square bond length $b$ counts the number of rings that need
to be expelled from the gyration volume of a given ring. This contribution
was conjectured \cite{Fischer} by an analysis of the number of concatenated
states in a melt of interpenetrating rings. The second term describes
the effect of topology in analogy to excluded volume \cite{Moore_Grosberg_2005},
which should be the best approximation for the weakly compressed rings
of the present study (cf. section \ref{sec:Internal-structure-of}
and ref \cite{Fischer}). The cross-over to strong compression at
$N_{C}$ is characterized by the transition to a dominating third
virial contribution $\sim b^{6}N^{3}/R^{6}$ that replaces the second
term in equation (\ref{eq:FET}) for $N>N_{C}$ as proposed by Grosberg
\cite{Grosberg_Feigel_Rabin}. A normalization constant $\mbox{v}_{e}$
for entangled strands is introduced in equation (\ref{eq:FET}) in
analogy to the excluded volume for entanglements. The equilibrium
ring size is, thus, obtained as 
\begin{equation}
R\sim N^{2/5}\left(\phi b/\mbox{v}_{e}\right)^{-1/5}.\label{eq:sc1}
\end{equation}
Above $N^{*}$, the number of rings that needs to be expelled is controlled
by the number of overlapping rings $\sim b^{3}N/R^{3}$, since $b\phi R^{2}$
cannot become larger than the number of overlapping rings. Compression
is in both cases balanced by a term $\sim b^{6}N^{3}/R^{6}$. This
yields ring sizes $R\sim bN^{3/8}$ for $N_{C}<N<N^{*}$ and $R\sim bN^{4/9}$
for $N>N^{*}$.

In the present work, we focus on bidisperse blends of ring molecules
at $N<N_{C}$ in order to clarify differences between previous theoretical
works \cite{Fischer,Sakaue2012,cates1986,Sakaue2011}. The idea behind
this procedure is that the majority species of rings determines the
properties of the melt. Therefore, inserting a small amount of minority
rings of different molecular weight into a melt of chemically identical
rings probes the properties of the surrounding melt and the topological
interactions between the rings. This is of particular importance,
since the models of Cates and Deutsch \cite{cates1986}, Sakaue \cite{Sakaue2012,Sakaue2011},
or Lang \emph{et al}. \cite{Fischer} implicitly or explicitly predict
different behaviours in bidisperse blends. While the first can only
be extended to have an onset of compression for melt molecular weight
$\sim N^{1/2}$, the second model requires the topological length
scale (entanglement length) to be decreasing for increasing molecular
weight. The third model assumes that the density of entanglements
is independent of $N$ for constant $\phi$ beyond the onset of the
effect of topology at the concatenation length $N_{OO}.$

In addition to the compression of rings in entangled melts of large
$b\phi R^{2}$ it is expected that the same principles as for linear
chains hold for the swelling of entangled rings in low molecular weight
rings. Let $M$ denote the degree of polymerization of the majority
species that dominates the melt properties and $N$ denote the degree
of polymerization of a chemically identical fraction of chains that
is dilute in the melt of $M$-mers. For linear chains, this situation
is typically discussed \cite{Degennes,Rubinstein} by using a Flory
ansatz of the form
\begin{equation}
\frac{\text{\ensuremath{\Delta}F}}{kT}\approx\frac{v}{M}\frac{N^{2}}{R^{3}}+\frac{R^{2}}{b^{2}N},\label{eq:FEEX}
\end{equation}
which balances the excluded volume contribution with the deformation
of an ideal chain. Here, $v\approx b^{3}$ is the excluded volume
per monomer. Equilibrium conformations are characterized by
\begin{equation}
R\sim bN^{3/5}M^{-1/5}.\label{eq:EQC}
\end{equation}
The onset of swelling is found by equating $N^{1/2}=N^{3/5}M^{-1/5}$,
which yields a matrix molecular weight $M_{c}\sim N^{1/2}$ to cross
over between swollen and nearly ideal conformations of the $N$-mers. 

In recent years, it was shown that bond-bond correlations introduce
corrections to the scaling of the size of linear chains \cite{wittmer2007intramolecular,Wittmer2004}.
Furthermore, one would expect that entanglement effects set in at
some degree of polymerization proportional to the entanglement degree
of polymerization $N_{e}$. Since $M_{c}$ is a function of $N$,
both regimes can overlap while simultaneously being perturbed by these
bond-bond correlations making a clean analysis quite problematic.
In order to circumvent such difficulties, two series of blends of
bidisperse rings were created which differ only by the fact that in
one series the topology of the rings is fixed because entanglements
are active (called ``entangled'' or ``non-interepenetrating''
samples), while for the other series entanglements are switched off
and excluded volume is maintained (``interpenetrating'' samples).
It is expected that the samples without entanglements show swelling
and related corrections in melt \cite{wittmer2007intramolecular,Wittmer2004}
similar to linear chains, a hypothesis that is successfully tested
below. In conclusion, a comparison between both cases then allows
for a rather direct determination of the effect of topology onto ring
conformations.

The paper is structured as follows: Section \ref{sec:Methods} describes
the simulation method and parameters of the simulations. Section \ref{sec:The-minimal-surface}
specifies the properties of the (minimal) surface that is spanned
by a ring polymer and the relation to the total number of rings in
topological conflict is explained. This analysis is fundamental for
understanding concatenation, since each concatenated ring must cross
the surface that is bounded by the ring. The internal structure of
the rings is compared with the available models in literatature \cite{Fischer,Sakaue2012,cates1986,Sakaue2011}
in section \ref{sec:Internal-structure-of} and it is shown that interpenetrating
rings show the same scaling and corrections due to excluded volume
as linear chains in melts. The sizes of interpenetrating and entangled
rings are compared in section \ref{sec:Dilute--mer-rings} and all
results are summarized in section \ref{sec:Discussion}.

\section{\label{sec:Methods}Simulation method and samples}

As in the preceding work \cite{Fischer}, we use the bond-fluctuation
model (BFM) \cite{CarmesinKremer88} to simulate bidisperse blends
of ring polymers. In this method, each monomer is represented by a
cube occupying eight lattice sites on a cubic lattice. In the standard
definition of this algorithm, the bonds between monomers are restricted
to a set of 108 bond vectors which ensure cut-avoidance of polymer
strands by checking for excluded volume. Monomer motion is modeled
by random jumps to one of the six nearest lattice positions. A move
is accepted, if the bonds connecting to the new position are still
within the set of bond vectors and if no monomers overlap. All samples
of the present study were created in simulation boxes of $128^{3}$
lattice sites with periodic bondary conditions at occupation density
$\phi=0.5$. 

In the present work, we use this method to create blends of non-concatenated
rings. Additionally, we perform simulations where we allow for an
extended set of bond vectors as described in \cite{Michalke2001}
such that all entanglements are switched off, while excluded volume
interactions are mainly unaffected. In consequence, the rings can
interpenetrate each other to form concatenated conformations. In both
cases, relaxation of the rings was monitored by the autocorrelation
function of the vectors connecting opposite monomers of a ring. All
samples were relaxed several relaxation times of the longest rings
in the bidisperse blends. Afterwards, chain conformations were analyzed
from snapshots of the ring solutions from a very long simulation run.
The error of the data points is computed by the total number of statistically
independent conformations available.

For the present study, we prepared bidisperse solutions of entangled
rings with a fraction of 1/32 of minority species with degree of polymerization
$N=2^{i}$ in melts of $M=2^{j}$ monomers, whereby $i$ and $j$
are all integers with $6\le i\le10$ and $4\le j\le10$. A second
series of interpenetrating rings with 1/32 fraction of $N=2^{i}$
and $6\le i\le9$ was prepared in the same range of melt degrees of
polymerization (except of skipping some of the largest $M$ for the
smallest two $N$ as indicated in the Figures). All samples containing
$N=1024$ were run for $10^{9}$ Monte Carlo Steps (``MCS'': attempted
moves per particle), the samples with $N=512$ for $5\times10^{8}$
MCS, and all other samples for at least $10^{8}$ MCS after equilibration.
The data below represent averages as determined over the full simulation
runs.

\section{\label{sec:The-minimal-surface}The minimal surface of a ring polymer}

\begin{figure}
\begin{center}\includegraphics[width=0.4\columnwidth]{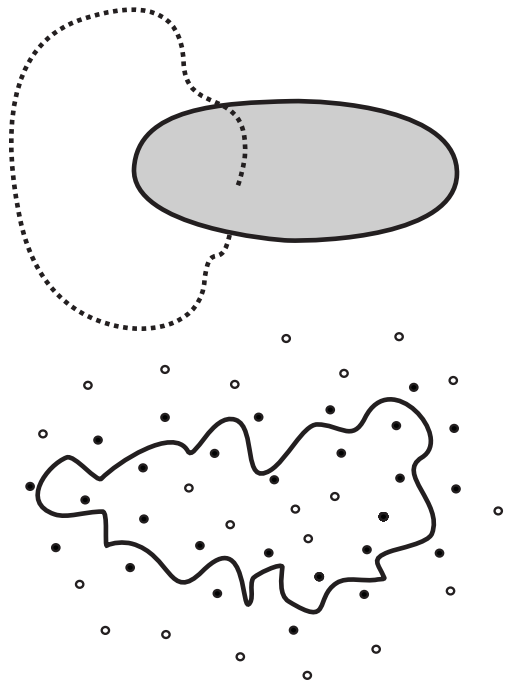}\end{center}

\caption{\label{fig:Upper-Figure:-concatenated}Upper Figure: concatenated
conformations (dashed line) intersect with the minimal surface of
the overlapping ring (grey color). Lower Figure: a different fraction
of overlapping strands is important for dynamics as compared to conformations:
the entanglements along the ring contour (full dots) inside and outside
of the ring for dynamics, and all entangled strands (full and open
dots) inside the ring contour for conformations.}
\end{figure}

Consider a circular random walk made by a randomly coiled wire. Dipping
this ring into a bowl of suds and removing it afterwards, one typically
observes the formation of a thin layer of suds held by the ring contour
that optimizes its area by surface tension. This simple experiment
determines the minimal surface bounded by the ring. The basic idea
of the model in reference \cite{Fischer} is that the (minimal) surface
spanned by the ring contour can be used to estimate the number of
topological conflicts among overlapping rings, since any concatenated
conformation has to pass through the area bounded by the ring, see
the upper part of Figure \ref{fig:Upper-Figure:-concatenated}. The
minimal property of this surface was originally chosen by convenience,
since concatenation can be detected by any surface bounded by the
ring. However, the minimal surface is the only uniquely defined surface
- to the knowledge of the author - that is in any possible situation
inside the volume spanned by the ring. Thus, crossing the minimal
surface is always equivalent to overlap \emph{and} possible concatenation,
which simplifies the discussion below. Therefore, the minimal surface
will be used in the present chapter to demonstrate that rings in melts
reduce the fraction of the bounded area that can be penetrated or
concatenated for avoiding concatenated states in melts of non-concatenated
rings. However, we have to keep in mind that the minimal surface is
only a concept to properly count or estimate the number of possible
concatenated states per ring but it is solely the number fraction
 of concatenated vs. not concatenated states (and not the minimal
surface) that drives the compression of non-concatenated rings in
melts.

In reference \cite{Fischer} it was assumed that a rather constant
fraction of all conformations that intersect with the minimal surface
produces a concatenated state. This assumption should be a reasonable
approximation as long as the root mean square lenght of the intersecting
walk is comparable or larger than the distance between intersection
point and boundary of the minimal surface. The observed good agreement
of the results for monodisperse melts of rings with the estimate based
upon the minimal surface corroborates this simplifying assumption.
But this simplification clearly deserves revision for mixtures of
rings with a broad distribution of ring sizes. Due to the limited
range of molecular weigths above $N_{OO}$ within the present study,
this discussion is not neccessary and thus, must be postponed to a
future work. In consequence of this simplification, we further can
assume an approximately constant concatenation probability as function
of the distance to the boundary of the minimal surface for the samples
of the present study. Within this approximation, the area distribution
of the minimal surface can be interpreted as distance distribution
of possible concatenations with respect to the boundary of the ring.

In polymer physics, the tube model is typically used to model the
effect of entanglements of overlapping polymer strands \cite{Edwards}.
In this model, it is assumed that the entanglements along the contour
of a chain confine the motion of the chain into a tube like region
in space and thus, predominantly the strands in contact with a given
molecule control the polymer dynamics. The situation is different
for the conformations of non-concatenated rings in melt. Any conformation
that would be concatenated needs to be expelled out of the volume
of a given ring, independent of the position of the concatenation
with respect to the ring contour. This difference is sketched in the
lower part of Figure \ref{fig:Upper-Figure:-concatenated}. Therefore,
any model (like the tube model) that neglects concatenations in the
inner part of the (minimal) surface is not suitable to learn about
ring conformations. Conversely, the topological approach of the present
work is also not suitable to learn about the dynamics of the rings,
because all entangled strands that are not part of concatenated conformations
are not detected in our analysis and thus, would not be taken into
account to estimate the dynamics of the ring.

To the knowledge of the author, the problem of the minimal surface
of a closed random walk is unsolved and it will not be attempted to
rigorously derive its properties. Knotted rings can be related to
Seifert-surfaces \cite{Adams}, but due to the fact that the knotting
length \cite{Fischer} is larger than the largest degree of polymerization
in our study, we ignore the effect of knotting for the remainder of
this work. Unknotted random walks bound much simpler surfaces: the
bounded area of an unknotted random walk is always homeomorph to a
circular disc \cite{Adams}. But the shape of this bounded area in
space is highly non-trivial.

\begin{figure}
\includegraphics[width=0.45\columnwidth]{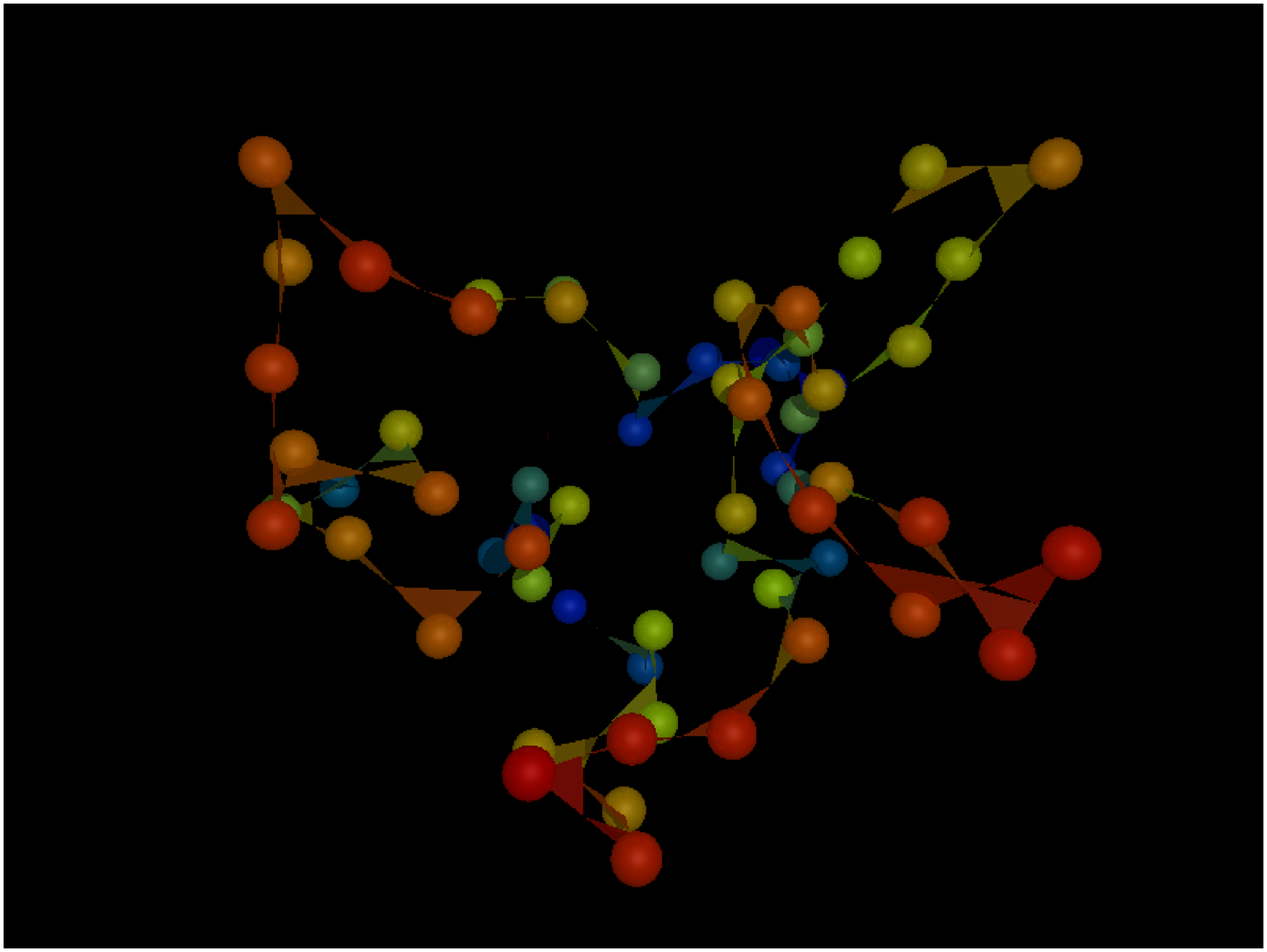}\includegraphics[width=0.45\columnwidth]{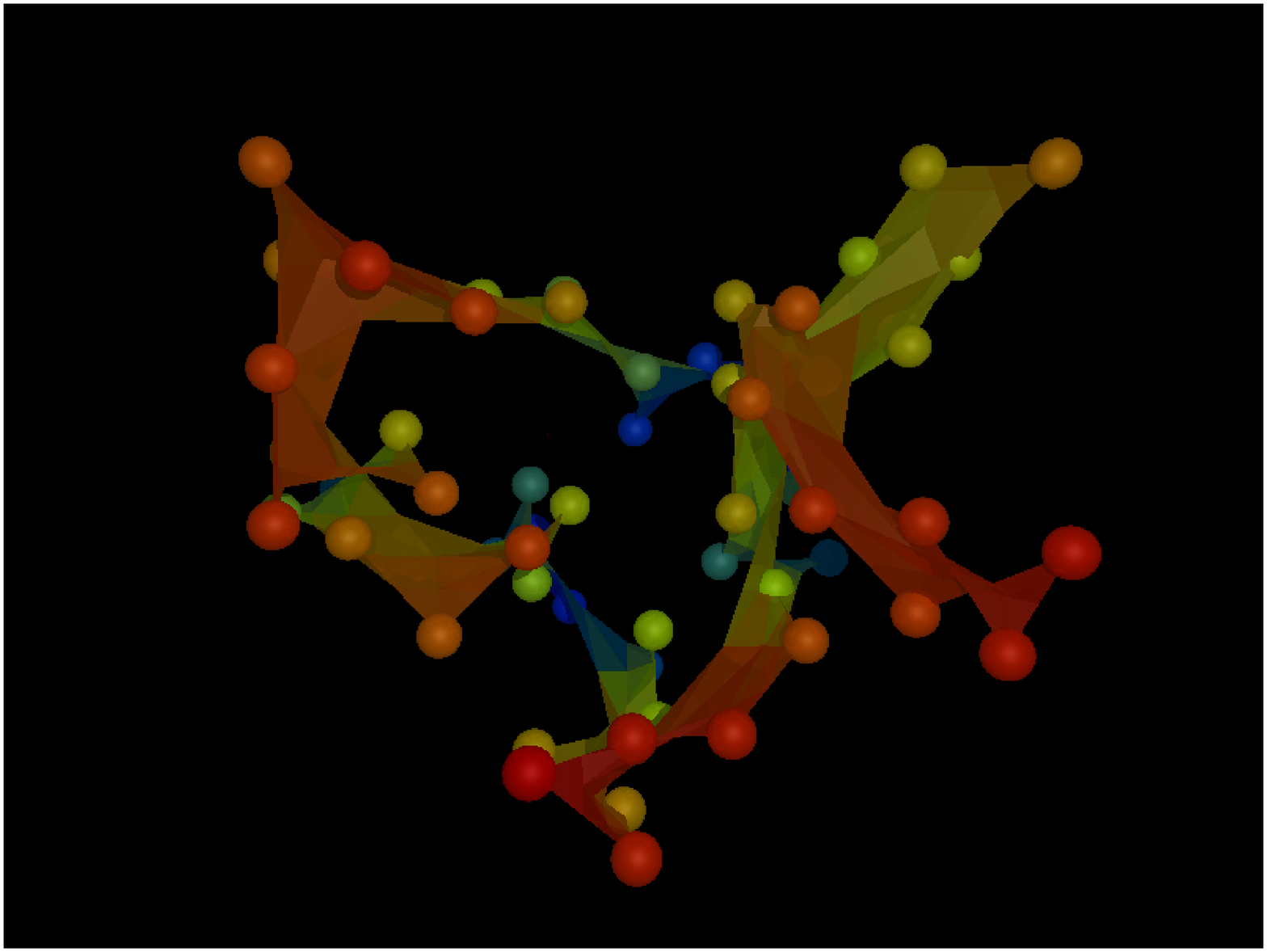}\vspace{-2mm}

\includegraphics[width=0.45\columnwidth]{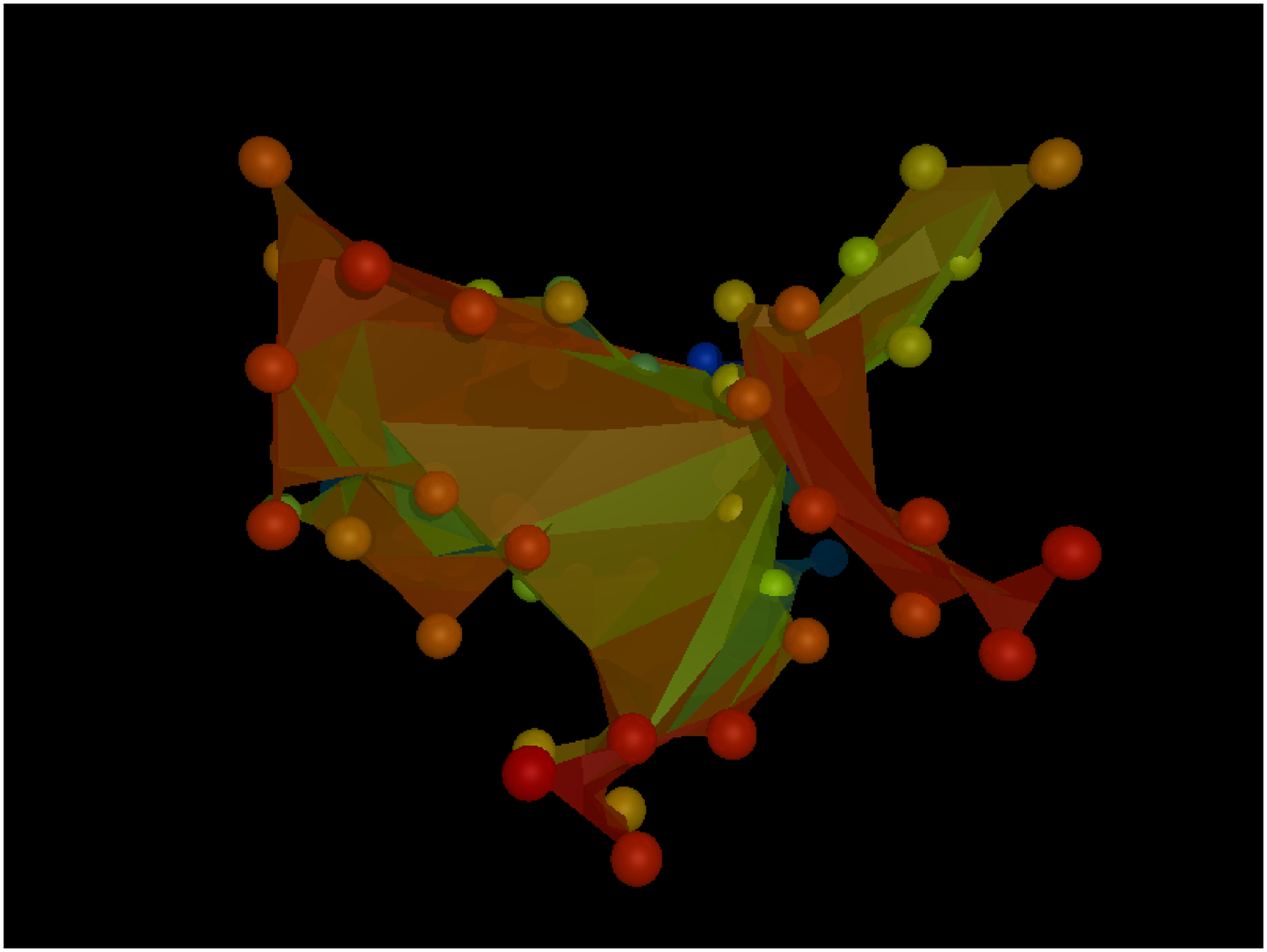}\includegraphics[width=0.45\columnwidth]{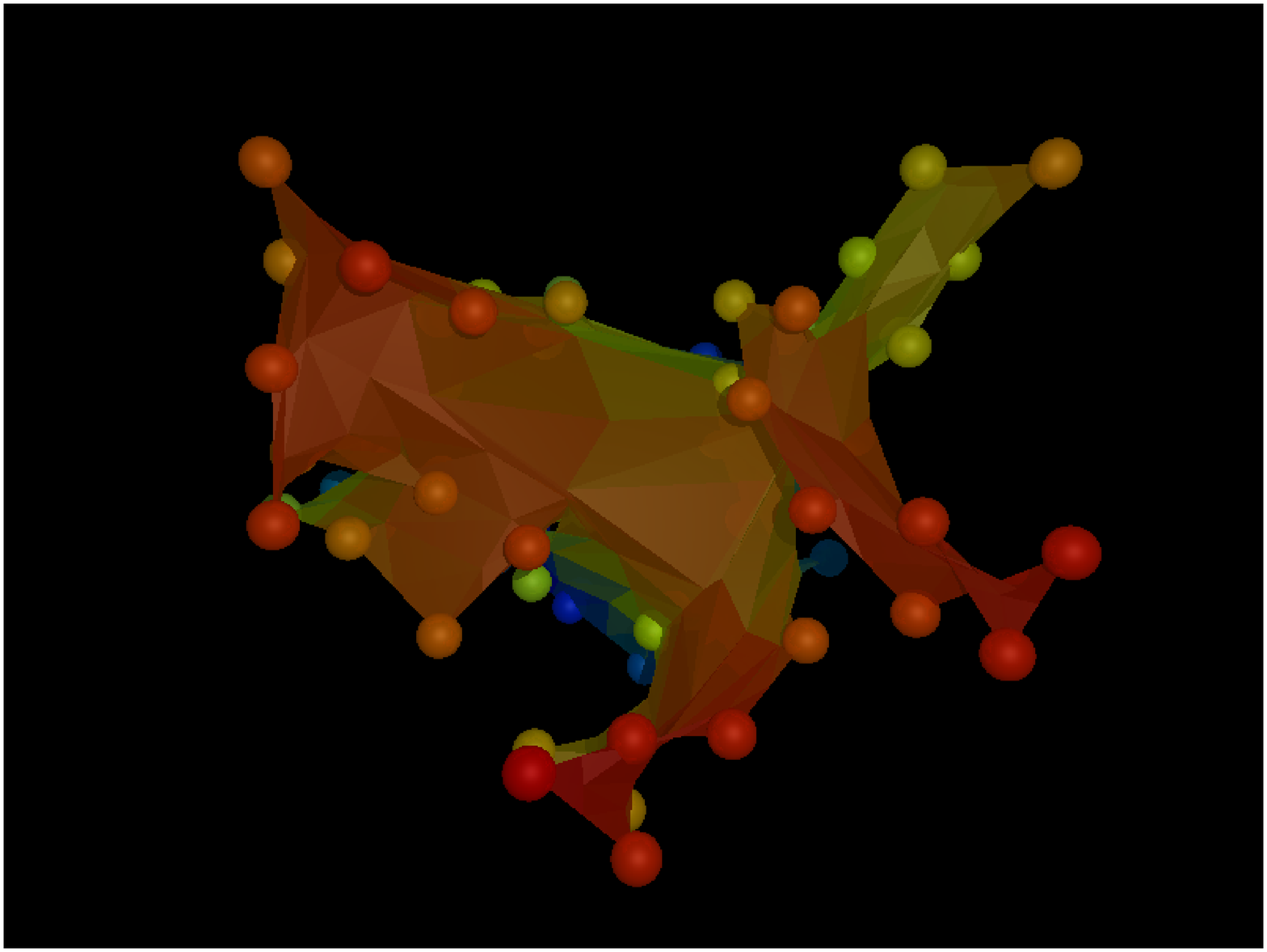}

\caption{\label{fig:Example}(All figures: color online) Left top: A (closed)
random walk and the first step of triangulation. Beads display monomers,
the color of the beads changes from red (closest bead) to blue (most
distant beads). Same color code applies for triangles and is determined
by the centers of mass of triangles. Bead size is $1/4$ of the size
in simulation to make the area along the boundary visible. Right top:
the first four generations of triangles (second generation of mobile
junctions) form a band along the ring contour of roughly $b$ width
with the inner area of the ring still uncovered. Left bottom: full
triangulation prior to optimization of area. Right bottom: fully optimized
area}
\end{figure}

In order to gain some insight into the properties of the minimal area
bounded by a ring polymer, an algorithm for the triangulation of the
area spanned by a ring was developed. Some steps of this triangulation
are shown in Figure \ref{fig:Example}. The total area was subdivided
into $6N$ triangles using $N$ joints along the ring and inserting
$N$ additional points at the middle of each segment. Further $2N-2$
mobile junctions are placed inside the area such that for each generation
of mobile junctions from boundary inwards the number of junctions
is divided by two. Each new junction is coupled to three junctions
of the previous generation and the next neighbors of the same generation.
Afterward, the total area of the surface is optimized by randomly
displacing one of the $2N-2$ inner mobile junctions using a Metropolis
\cite{Metropolis} algorithm for the change in area while quickly
driving the ``temperature'' to zero. Optimization was stopped, if
five consecutive cycles over all mobile junctions did not further
minimize the area. This rapid cooling protocol was tested against
a slow minimization and shown to lead to an inaccuracy of approximately
$1\%$ as compared to the best results. Note that such a rapid cooling
scheme and this rather coarse triangle mesh were necessary to be able
to analyze all available ring conformations. After minimization, the
area of each triangle was computed and the area analyzed as function
of the distance between the center of mass of the triangle and the
nearest ring monomer. The results are discussed in the following paragraphs.

\begin{figure}
\includegraphics[angle=270,width=0.8\columnwidth]{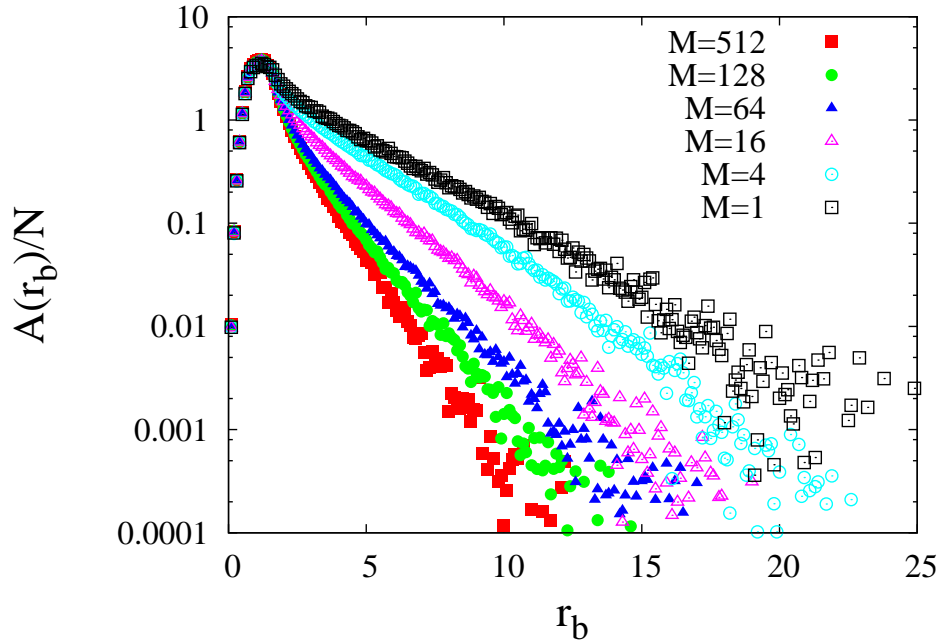}

\caption{\label{fig:Area--per}Area $A$ per bond as function of the minimum
distance $r_{b}$ to the boundary for a series of ring polymers with
$N=512$ monomers in melts of $M$ monomers. The length unit is the
length unit of the simulation lattice as in all Figures below.}
\end{figure}

\begin{figure}
\includegraphics[angle=270,width=0.8\columnwidth]{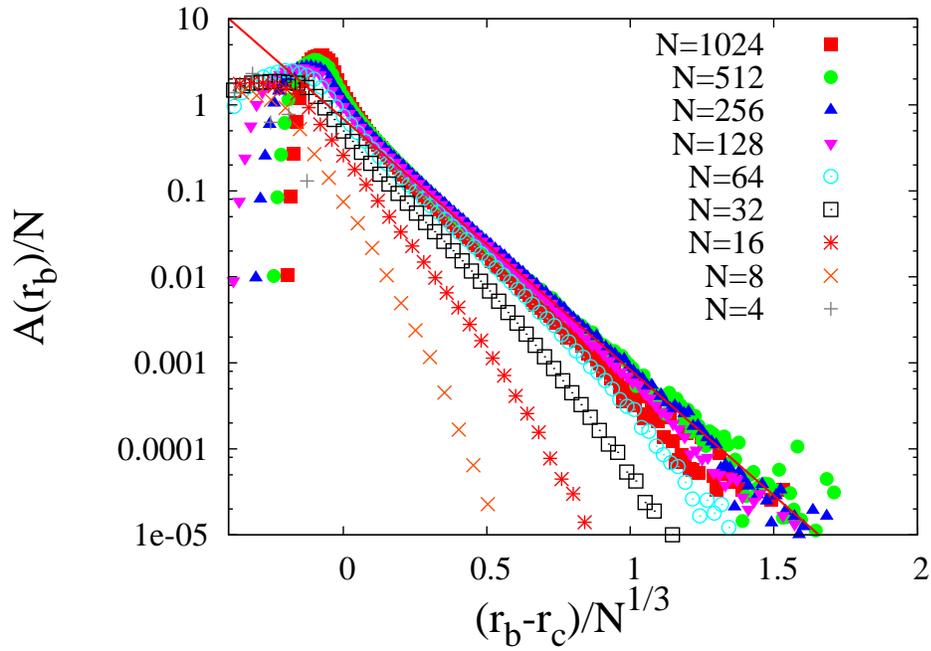}

\caption{\label{fig:Area--per-2}Area $A$ per bond as function of the minimum
distance $r_{b}$ to the boundary for a series of monodisperse, entangled
rings in melt. }
\end{figure}

The results for a series of entangled samples with long $N$-mers
of 512 monomers in melts of rings of $M$ monomers is shown in Figure
\ref{fig:Area--per-2}. Here, $r_{b}$ is the minimum distance of
the center of mass of an area segment to the nearest monomer of the
ring. All data show a peak at $r_{b}\le2$ and an exponential decreasing
area as function of $r_{b}>2$. Note that a monomer in the simulations
has a diameter of 2 lattice units and thus, the minimum distance of
area sections that can be penetrated is $r_{c}=2$. The data overlap
in the region of the peak, while the exponential decrease depends
clearly on the molecular weight of the surrounding melt. Note that
the samples $M<16$ refer to swollen rings and $M>16$ to compressed
rings, while $M=16$ is close to being ideal (up to corrections to
screening, as discussed later). Figure \ref{fig:Area--per-2} clearly
demonstrates that rings change their conformations upon swelling or
compression such that mainly area sections with large distance to
the boundary are created or destroyed. The area along the boundary
remains mainly unmodified.

\begin{figure}
\includegraphics[angle=270,width=0.8\columnwidth]{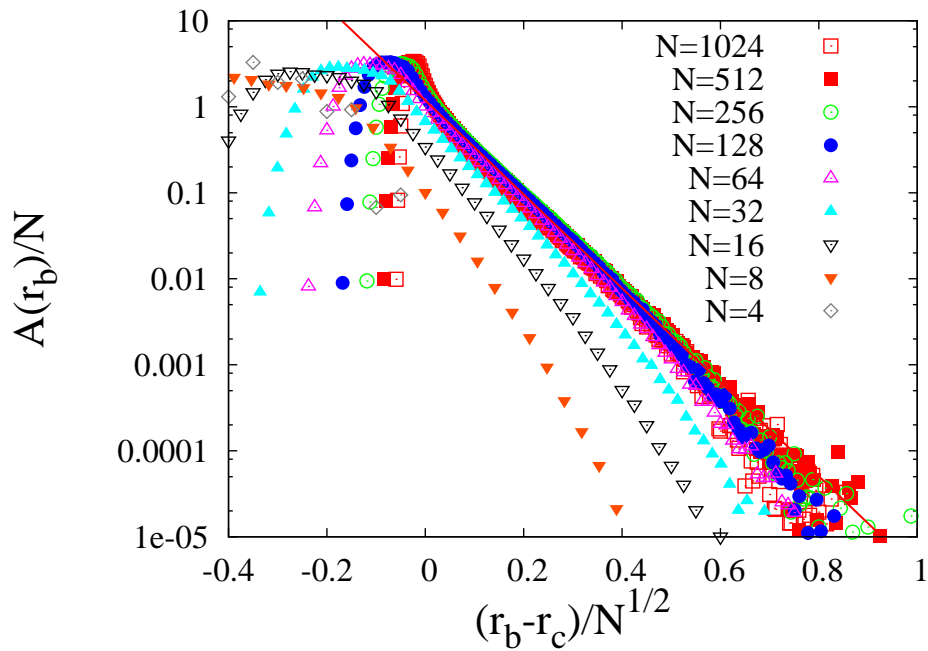}

\caption{\label{fig:Area--per-1}Area $A$ per bond as function of the minimum
distance $r_{b}$ to the boundary for a series of freely interpenetrating
monodisperse melts of rings of $N$ monomers. The line indicates an
exponential decay.}
\end{figure}

\begin{figure}
\includegraphics[angle=270,width=0.8\columnwidth]{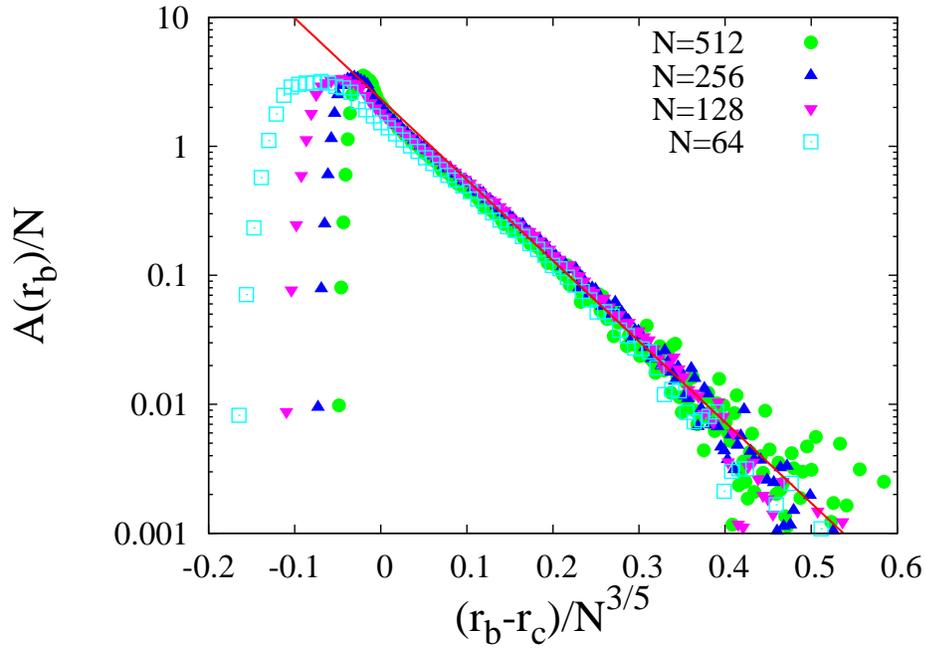}

\caption{\label{fig:Area--per-1-1}Area $A$ per bond as function of the minimum
distance $r_{b}$ to the boundary for a series of fully swollen non-knotted
non-concatenated rings in monomeric solvent.}
\end{figure}

In order to understand the exponential tail of the area distribution
it is attempted to search for scaling variables that lead to a collapse
of the data. The simplest expectation is to assume that the distance
to the central sections of the minimal surface might scale as the
diameter of the ring, which is $\sim N^{1/2}$ for the freely interpenetrating
rings. In order to test this hypothesis, the area distribution of
a series of freely interpenetrating ring melts is plotted in Figure
\ref{fig:Area--per-1} vs $(r_{b}-r_{c})/N^{1/2}$, which collapses
the data for sufficiently large $N$. The same collapse is obtained
for the data of sufficiently large swollen rings in monomeric solvent
when using the same cut-off $r_{c}$ and the corresponding power of
$\nu\approx0.588$ \cite{Clisby}, as demonstrated in Figure \ref{fig:Area--per-1-1}.

Figure \ref{fig:Area--per-2} shows that the ring area of entangled
melts of rings is reduced quicker than the total size of the ring:
instead of for $N^{2/5}$, a best overlap of the data at large $N$
is found for $(r_{b}-r_{c})/N^{1/3}$. This difference suggests that
a reduction in area for preventing penetration might drive the compression
of the rings. The total area for penetration is of order $N^{4/3}$
and thus, still an increasing function of $N$. Therefore, penetrations
of double folded sections of overlapping rings are clearly reduced
as compared to interpenetrating melts, but are not fully excluded
for large $N$. This is a crucial point for understanding the dynamics
of melts of rings and deserves a more detailed investigation in a
forthcomming work.

Figures \ref{fig:Area--per} to \ref{fig:Area--per-2} suggest that
a cut-off of $r_{c}\approx2$ at $\phi=0.5$ defines the onset of
the exponential tail of the area distribution%
\footnote{Note that $r_{c}$ should scale as the blob size in semi-dilute solutions
of rings.%
}. The area very close to the boundary $r_{b}<1$ cannot be distinguished
from the effect of excluded volume and is not accessible to neither
bonds nor monomers nor it can be adjusted by creating double folds.
Therefore, the area at $r_{b}<1$ is expected to remain constant as
function of $N$. The area at distance $1<r_{b}\lesssim2$ lattice
units is subject to local conformational changes or self contacts
but not to penetration by other rings. Penetration requires a minimum
area diameter of $\approx4$ lattice units, since the monomer diameter
is two lattice units.

The exponential tail $\sim\exp\left(-a\left(r_{b}-r_{c}\right)/N^{1/2}\right)$
in Figure \ref{fig:Area--per} at $r_{b}>2$ implies a correction
$\sim N^{1/2}$ to scaling leading a total area of order $R^{2}N^{1/2}$.
The number of monomers of an overlapping ring of same degree of polymerization
$N$ in contact with the minimal surface is of order $\sim N^{1/2}$.
Thus, the number of rings that have to be expelled is $f_{n}\sim R^{2}$,
as found previously in ref. \cite{Fischer}. However, there is a more
subtle point missing in this discussion: an inner area $\sim R^{2}N^{1/2}$
is equivalent to a volume of order $R^{3}$ due to $R\sim N^{1/2}$
for random walks. Furthermore, a scaling of the inner area $\sim R^{2}N^{1/2}$
is only possible, if the average curvature of the minimal surface
is rather independent of $N$. The point is that sufficently small
sections of the inner area appear locally flat as long as the size
of the area is below the curvature radius of the minimal surface.
Larger areas fully cover space and the number of returns of a second
random walk to this curved area becomes $\sim N$. If this is the
case, the total number of excluded rings becomes $\sim R^{2}N^{-1/2}$,
which is $\sim N^{1/2}$ and, thus, proportional to the overlap number.
According to the fraction of overlapping pairs of rings that mutually
would entangle each other, this is expected in the limit of very large
$N$ above a critical degree of polymerization $N^{*}$ \cite{Fischer}
that is beyond the largest degree of polymerization in the present
study.Finally, since semi-dilute solutions of polymers can be described
by a melt of blobs, rescaling $N$ by the number $g$ of monomers
in a blob brings us back to the above estimate in melts, since both
bounding rings and penetrating rings are rescaled by the same $g.$Altogether,
the inner part of the area leads to a penalty for non-concatenation
$\sim\phi R^{2}$ as introduced in ref \cite{Fischer}, since the
density of surrounding rings is $\sim\phi$.

\begin{figure}
\includegraphics[angle=270,width=0.8\columnwidth]{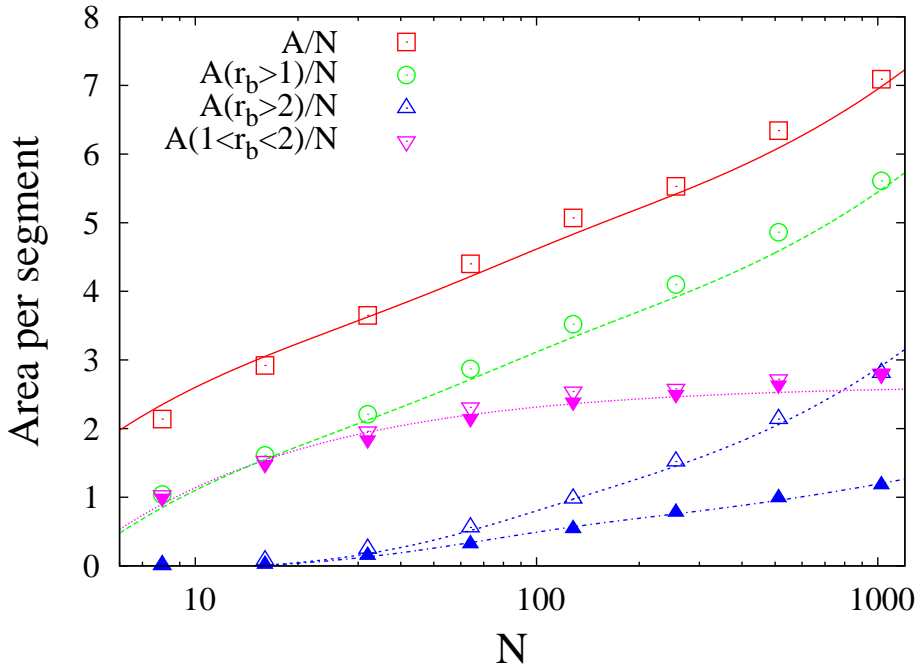}

\caption{\label{fig:Total-area-per}Total area per bond $A/N$ (measured in
square lattice units) and the area fraction $A(r_{b}>1)/N$ with its
contributions close to the boundary $A(1<r_{b}<2)/N$ and further
away from the boundary $A(r_{b}>2)/N$ in comparison. Open symbols
is data of interpenetrating rings, full symbols additional data of
entangled rings.}
\end{figure}

The resulting picture for the minimization of the bounded area for
$N\le1024$ of the present study can be discussed by using the data
of Figure \ref{fig:Total-area-per}. The lines in Figure \ref{fig:Total-area-per}
serve merely as a test of consistency of the above analysis. Note
that a test with a limited number of rings but a finer mesh for triangulation
showed that the absolute contribution of the different sections of
the area are slightly modified with increasing number of triangles,
but the function describing the form of the contributions is merely
unchanged. First, the fraction of the area at $r_{b}<1$ remains mainly
constant, as explained above. Up to $N\approx100$, the fraction of
area close to the boundary $A(1<r_{b}<2)$ that is subject to self-contacts
and local unfolding of rings is increasing in Figure \ref{fig:Total-area-per}
and getting approximately constant for large $N$. As mentioned above,
this is mainly because local sections of rings unfold and become more
and more similar to the same section of a linear chain for increasing
$N$. The fraction of the inner area $A(r_{b}>2)$ in Figure \ref{fig:Total-area-per}
fits well against $\sim N^{1/2}(1-P_{OO})$ using the non-concatenation
probability $P_{OO}=\exp\left(-(N-a)/N_{OO}\right)$ with $a=15.0\pm0.5$
and $N_{OO}=40.0\pm0.2$ at $\phi=0.5$, as described in ref. \cite{Fischer}.
Thus, the data on the exponential tail and the data on the number
of linked rings $f_{n}$ of ref. \cite{Fischer} agree well with each
other. The fact that the compression of rings is only driven by the
changes in the inner area is demonstrated by comparing with the inner
area of entangled rings: the area fractions at $r_{b}<2$ are the
same in both cases, which is exemplified here by comparing with the
non-constant fraction $A(1<r_{b}<2)$, but the inner part of the area
is clearly reduced following a dependence $\sim N^{1/3}(1-P_{OO})$. 

It is quite interesting to note that even for the largest $N$ the
inner fraction of the area is still not the largest part of the minimal
area. When extrapolating the data, inner area and boundary become
comparable in the range of the knotting length $N_{0}$ of these samples
\cite{Fischer}. Knotting as well as linking other rings requires
an open area that can be penetrated by a strand of the same polymer
(knotting) or different polymer (linking). Therefore, it might be
an interesting subject for a future study to compare knotting of individual
rings of variable excluded volume with the change in the inner minimal
area of ring polymers.

\section{\label{sec:Internal-structure-of}Internal structure of ring polymers
in a melt}

The models of Lang \emph{et al. }\cite{Fischer}, Cates and Deutsch
\cite{cates1986}, and Sakaue \cite{Sakaue2012,Sakaue2011} differ
clearly in their predictions or assumptions on the microscopic structure
of rings in a melt of rings. Sakaue \cite{Sakaue2012,Sakaue2011}
postulates a change in the topological length scale as function of
melt molecular weight. Thus, local conformations of rings are expected
to be a function of melt molecular weight. In particular, the onset
of compression is expected to move to smaller distances in space and
along the chains with increasing $M$.

Cates and Deutsch do not discuss the bidisperse case, however, a generalization
of scaling to bidisperse blends is straight forward. The size of swollen
rings in good solvent is fixed at $R_{s}\sim M^{-1/5}N^{3/5}$ and
in monodisperse melts $R_{m}\sim N^{2/5}$. We further note that in
the approach of Cates and Deutsch and, thus, in equation (\ref{eq:FEEX}),
only $R,$ $N$, and $M$ enter. Therefore, there are no other physical
length scales in this model and any cross-over can only be a function
of these variables. Searching for the cross-over point we find $M_{c}=N$,
which is also the limiting case under which such a scaling could be
applied. Thus, the only possible scaling of ring conformations in
bidisperse melts consistent with the Cates and Deutsch conjecture
is 
\begin{equation}
R\sim M^{-1/5}N^{3/5}\label{eq:asdf}
\end{equation}
for $M\le N$ and 
\begin{equation}
R\sim N^{2/5}\label{eq:sdf}
\end{equation}
for $M>N$. On a microscopic level, this is consistent with rings
that are compressed on local scale comparable to the size of the surrounding
rings while staying ideal on larger length scales. Furthermore, rings
are expected to be ideal for $M_{c}=N^{1/2}$ and thus, the onset
of compression is expected to be a function of $N$ independent of
the entanglement degree of polymerization $N_{e}$ of the polymers.

The model of Lang \emph{et al. }\cite{Fischer} makes the following
assumptions or predictions: a) the compression of rings in melt is
related to the number of overlapping rings that would be mutually
concatenated. b) The topological length scale (entanglement length)
is constant. The onset of compression is controlled by the onset of
mutual penetration of rings at $N_{OO}$. c) Since the rings adopt
conformations that are subject to minimizing the bounded area, a significant
increase of the fraction double folded ring sections can be expected.
These folds could become visible as an anti-correlation peak in the
bond-bond correlation function.

The different predictions or assumptions of the models above are now
tested in detail. Let us first treat ideal rings of $N$ monomers
as basis for our discussion. Let $1\le S\le N/2$ denote the minimum
number of segments between monomer $i$ and $j$ along the ring. The
two strands of $S$ and $N-S$ connecting monomer $i$ and $j$ act
like an elastic chain of length 
\begin{equation}
s=\frac{S(N-S)}{N}\label{eq:s}
\end{equation}
connecting both monomers. The average distance between monomer $i$
and $j$ in an ideal ring is therefore 
\begin{equation}
R_{e}(s)=bs^{1/2}.\label{eq:Re}
\end{equation}

\begin{figure}
\includegraphics[angle=270,width=0.8\columnwidth]{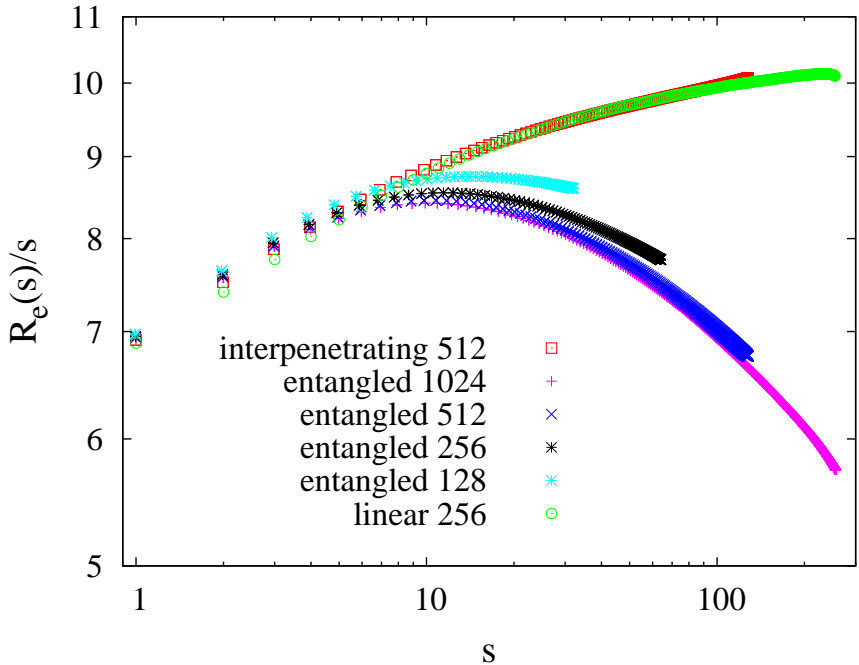}

\caption{\label{fig:Normalized-mean-square}Normalized mean square distance
between monomers separated by an elastic strand of $s$ segments in
monodisperse melts.}
\end{figure}

Wittmer et al. \cite{wittmer2007intramolecular,Wittmer2004} reported
corrections to the ideal behaviour for linear chains in melt, which
result in long range bond correlations and a slow convergence of chain
size towards the asymptotic limit. This slow convergence is visible
as correction to the average root mean square distance between to
monomers along a linear chain. As shown in Figure \ref{fig:Normalized-mean-square},
the smallest $s$ contribute most of this correction for linear chains.
Obviously, $s=S$ for linear chains. For long rings $N\gg1$ there
is $s\approx S$ at small $S$. Therefore, we do not expect large
deviations between long linear chains in melts and long interpenetrating
rings that are not affected by entanglements. This is demonstrated
by the excellent overlap of the data of interepenetrating rings and
linear chains in Figure \ref{fig:Normalized-mean-square} when plotting
$R_{e}(s)$ as function of $s$ both times. This overlap further shows
that our samples were well equilibrated and the low scatter of the
data (which is in fact the correction to scaling of ring size) indicates
the high precision of the results. Note that switching off entanglements
(see section \ref{sec:Methods}) increases the average square bond
length from 6.95 to 7.45 due to the modification of the set of bond
vectors. Overlap of the data was thus obtained by renormalizing size
by the change in the average square bond length of the interpenetrating
rings.

The data of the entangled ring samples shows a clearly different behaviour
in Figure \ref{fig:Normalized-mean-square}, whereby the sample with
$N=128$ is at the onset of the fully entangled regime (cf. below).
For larger rings, at $s\approx10$, the tangent to the mean square
distance becomes horizontal indicating that the remaining effect of
excluded volume is fully compensated by topological interactions.
For larger $s$, rings become increasingly compressed with increasing
$s$, whereby the data of large $N$ appear to fall on a rather universal
curve. The short distance behaviour is nearly unaffected. This observation
is clearly different to the model of Sakaue \cite{Sakaue2012,Sakaue2011},
which postulates a change in the topological length scale as function
of melt molecular weight. A change in the topological length scale
should lead to a shift of the peak position as function of $s$. Such
a behaviour cannot be deduced from Figure \ref{fig:Normalized-mean-square}.
Similarily, the extended model of Cates and Deutsch would also predict
a shift of the onset of compression as function of $M\sim N^{1/2}$.
This is not supported by Figure \ref{fig:Normalized-mean-square}. 

It is worth pointing out that even though the rings are compressed
as compared to their linear counterparts, this compression is still
so small that rings up to $N\approx500$ monomers are slightly swollen
in Figure \ref{fig:Normalized-mean-square} as compared to ideal rings.
This can be seen from $R_{e}(s)/s>R_{e}(1)$, since for ideal rings
$R_{e}(s)/s\equiv R_{e}(1)$.

\begin{figure}
\includegraphics[angle=270,width=0.8\columnwidth]{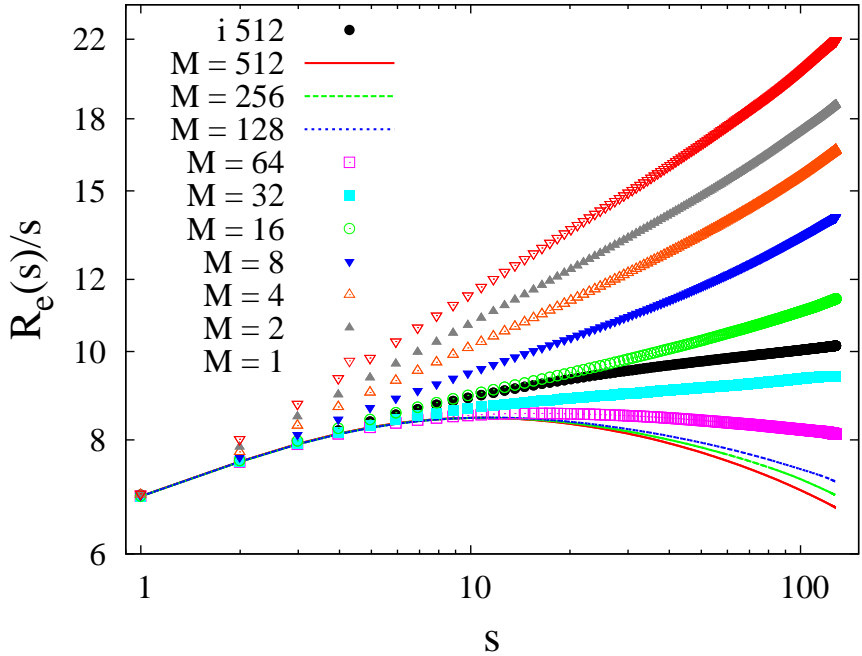}

\caption{\label{fig:Normalized-mean-square-1}Normalized mean square distance
between monomers separated by an elastic strand of $s$ segments.
Data of entangled rings of $N=512$ are shown as function of the degree
of polymerization $M$ of the surrounding melt of entangled rings
(and in melts of short linear chains for $M\le2$). The monodisperse
sample $N=512$ of interpenetrating rings (sample i 512) is included
for comparison.}
\end{figure}

The internal distances inside the rings are analyzed as function of
melt molecular weight, see Figure \ref{fig:Normalized-mean-square-1}.
Again, the deviation from the monodisperse case $M=512$ starts at
large distances, whereby short distances are nearly unaffected. For
$M\le16$, which is equivalent to $M<N^{1/2}$, the ring conformations
are additionally affected by swelling (as compared to the monodisperse
reference sample $N=512$ of interpenetrating rings) due to low molecular
weight solvent. This type of swelling manifests itself in parallel
sets of data at large $s$ for decreasing $M$, since the onset of
swelling further propagates to smaller $s$. The compression of rings
with $M\ge32$ is qualitatively different, since the data at large
$s$ is no longer parallel.

Wittmer et al. \cite{wittmer2007intramolecular,Wittmer2004} identified
time average bond-bond correlations $\left\langle \mathbf{b}_{j},\mathbf{b}_{i}\right\rangle $
as source of changes in the internal distances $R_{e}(s)$. Let the
vector $\mathbf{r}_{j}$ denote the coordinates of monomer $j$ and
$\mathbf{b}_{j}=\mathbf{r}_{j+1}-\mathbf{r}_{j}$ the bond vector
connecting monomer $j$ and $j+1$. Note that rings are ``periodic''
in the sense that $\mathbf{r}_{N+i}=\mathbf{r}_{i}$. Furthermore,
ring closure is expressed by $\sum_{i=1}^{N}\mathbf{b}_{i}=\mathbf{0}.$
This condition implies $\sum_{i}\left\langle \mathbf{b}_{j},\mathbf{b}_{i}\right\rangle =0$
and $\mathbf{b}_{i}=-\sum_{j\ne i}\mathbf{b}_{j}$. Due to $\left\langle \mathbf{b}_{i},\mathbf{b}_{i}\right\rangle =b^{2}$
for any $i$ with root mean square bond length $b$ we thus have the
average correlation 
\begin{equation}
\left\langle \mathbf{b}_{j},\mathbf{b}_{i}\right\rangle =-b^{2}/(N-1)\label{eq:c2}
\end{equation}
among two bonds $i\ne j$ in an ideal ring.

\begin{figure}
\includegraphics[angle=270,width=0.8\columnwidth]{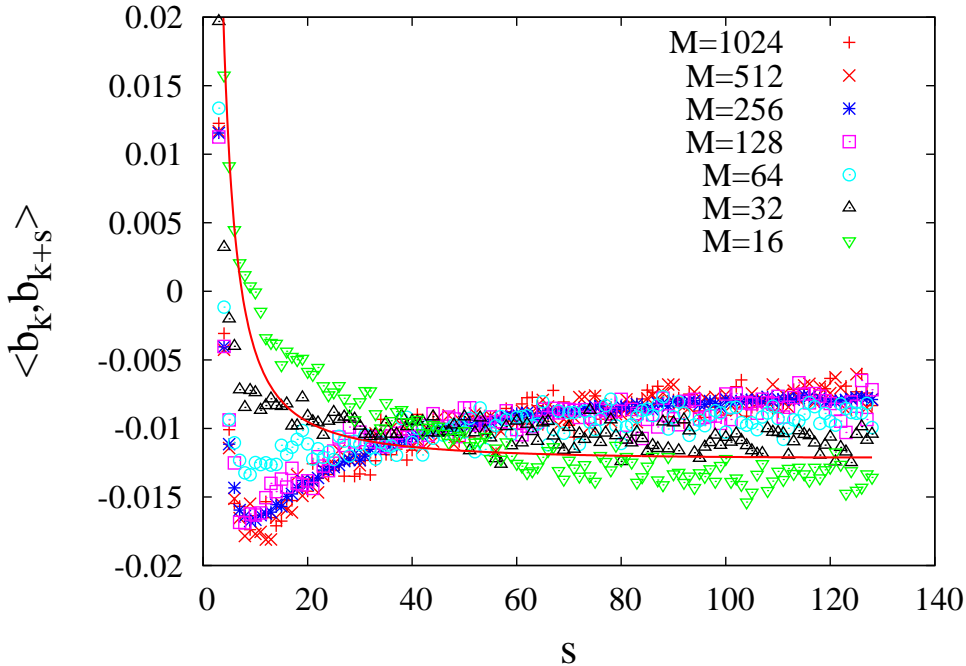}

\caption{\label{fig:Bond-correlations-in-1}Bond correlations in entangled
melts of rings for $N=256$. The line is the same approximation as
used for the unentangled data Figure \ref{fig:Bond-correlations-in-1}
for comparison.}
\end{figure}

Excluded volume interactions introduce long range bond-bond correlations
that decay as a power law \cite{wittmer2007intramolecular,Wittmer2004}.
These correlations lead to a renormalization of the effective bond
length in melt, such that the chain conformations can be described
by a random walk with an effective bond length $b_{e}\approx1.23b$
at $\phi=0.5$ in the limit of large $N$. For very large rings with
excluded volume we, thus, expect that the bond-bond correlation function
$\left\langle \mathbf{b}_{j},\mathbf{b}_{j+S}\right\rangle $ drops
from the linear chain power law $S^{-3/2}$ in a melt of long rings
at small $S$ to approximately $\left\langle \mathbf{b}_{j},\mathbf{b}_{j+S}\right\rangle \approx-b_{e}^{2}/(N-1)$
for $S\approx N/2$. Fitting the monodisperse melt data of $M=256$
with $\left\langle \mathbf{b}_{j},\mathbf{b}_{j+S}\right\rangle =const\cdot S^{-3/2}-b_{e}^{2}/(N-1)$
in Figure \ref{fig:Bond-correlations-in-1} (black line) yields reasonable
agreement supporting this simple approximate treatment. The data further
show that bond correlations become universal in melts $M\gtrsim N^{1/2}$.

\begin{figure}
\includegraphics[angle=270,width=0.8\columnwidth]{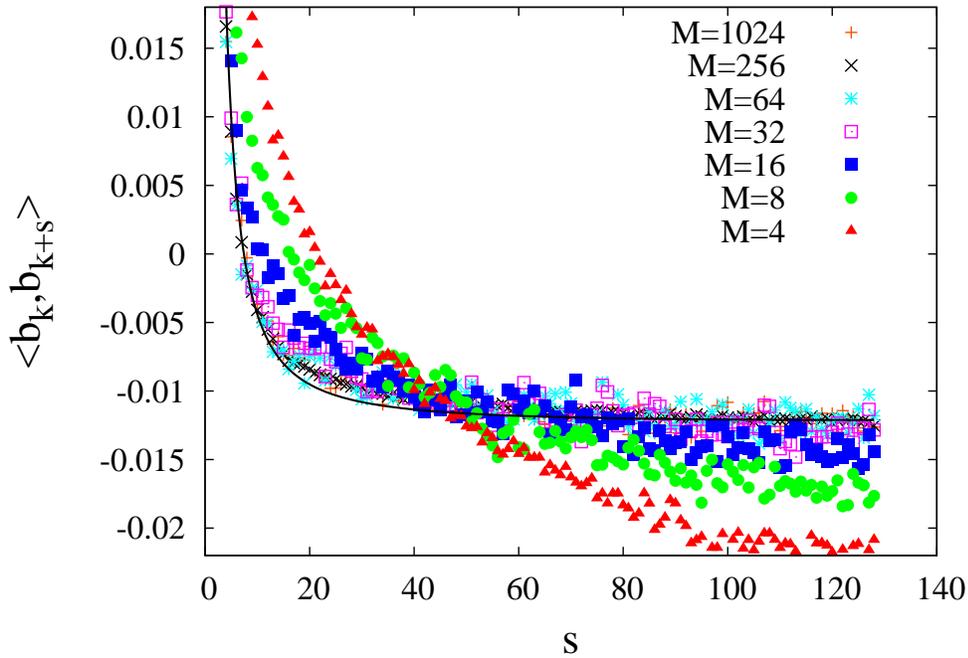}

\caption{\label{fig:Bond-correlations-in}Bond correlations in a melt of interpenetrating
rings of $M$ monomers as experienced by rings with $N=256$. The
line is a fit to the data $M=256$ as described in the text.}
\end{figure}

The same analysis for the entangled samples yields a clearly different
result. For sufficiently large $N$, an anti-correlation peak is formed
around $S\approx10$ in Figure \ref{fig:Bond-correlations-in-1} for
increasing melt degree of polymerization $M$. A similar anti-correlation
has been observed previously in monodisperse melts of rings \cite{MuellerII}.
Peak position and amplitude become independent of melt molecular weight
for $M\ge128$ similar to the position of the maximum in Figure \ref{fig:Normalized-mean-square-1}.
The anti-correlation peak also leads to a reduction of anti-correlation
at large $S$.

Both observations (anti-correlation and maximum in normalized internal
distances) could be explained, if the rings contain a significant
amount of short double folded sections of about 10 monomers. The formation
of such double folds, however, requires that topology is important
for short strands (of long rings) at about 10 monomers. Note that
an onset of entanglement effects at the same $s\approx10$ was found
in a recent work on networks \cite{Lang2010}. As mentioned above,
Sakaue \cite{Sakaue2012,Sakaue2011} postulates a shift of the topological
length scale as function of $M$. Also, the CD model requires the
onset of compression to be $\sim M^{1/2}$. This is in clear disagreement
to the constant peak position in Figure \ref{fig:Bond-correlations-in-1}
for sufficiently large $M$.

\section{\label{sec:Dilute--mer-rings}Size of dilute $N$-mer rings in melts
of $M$-mer rings: a comparison of entangled and interpenetrating
rings}

In order to subtract the effect of excluded volume swelling from ring
compression, the ratio of the sizes of rings in the interpenetrating
and the entangled samples, $R_{in}/R_{en}$, is analyzed. The result
is taken to power $5/2$, since then, all corrections to the first
term of the free energy in equation (\ref{eq:FET}) show up in first
power. In ref \cite{Fischer} it was shown that the predominant correction
at small $N$ should be a cut-off for non-concatenation of form $P_{OO}=\exp\left(-(N-a)/N_{OO}\right)$.
This cut-off was determined in monodisperse melts. For an $N$-mer
ring in a melt of $M$-mer rings, the cut-off is controlled by the
$M$-mer rings, if $N>M$. Therefore, we have to plot the data as
function of $M$ and use a function of form $P_{OO}(M)=\exp\left(-(M-a^{*})/N_{OO}\right)$
to describe the onset of concatenation, and thus, compression of the
$N$-mer. No further changes in ring size are expected in melts of
sufficiently large $M>N_{OO}$ beyond the cut-off $N_{OO}$ and the
cut-off $N_{OO}$ itself should remain unaffected for $N>N_{OO}$.
As compared to the monodisperse case, the onset of compression as
modeled by parameter $a$ might be reduced for large $N$-mers in
a melt of $M$-mers, since now only the smaller $M$-mer requires
to locally unfold to allow for penetration. Therefore, $a^{*}$ was
introduced as adjustable parameter and it was found that the data
at small $M$ is best approximated for an $a^{*}\approx a/2$ when
being compared with the monodisperse $a$.

\begin{figure}
\includegraphics[angle=270,width=0.8\columnwidth]{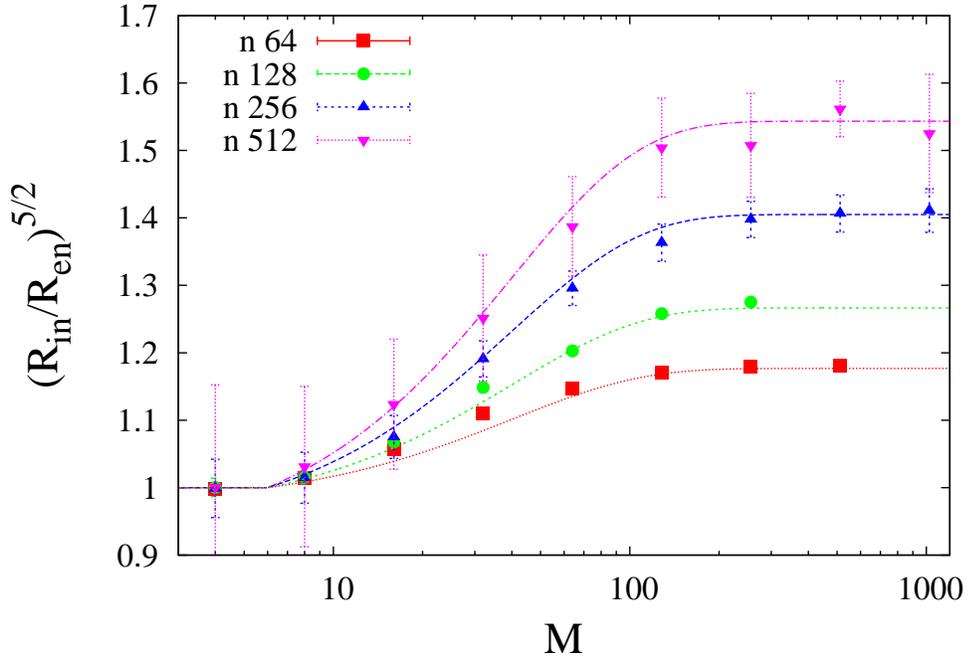}

\caption{\label{fig:Size-of-non-concatenated}Size of interpenetrating rings
normalized by the size of entangled rings.}
\end{figure}

Figure \ref{fig:Size-of-non-concatenated} compares the $\left(R_{in}/R_{en}\right)^{5/2}$
and the onset of compression towards larger $M$ with a function of
form $const\cdot\left(1-P_{OO}\right)+1$. The good agreement of the
onset of compression with this function shows that the correction
$P_{OO}$ is the dominating correction for the onset of ring compression
in bidisperse blends. The plateau at large $M$ demonstrates that
ring size is approximately constant in melts of larger degree of polymerization,
at least for the range of $100\lesssim N\lesssim1000$, as accessible
in the present study. This is because the eponential decay of the
distribution of inner areas suppresses further corrections as function
of $M$ for the rather small ratio of $N/M$ available with $N,M>N_{OO}$
and $N,M\lesssim1000$. Obviously, the level of the plateau for compression
at $N\ge128$ is in agreement with a scaling of ring size $\sim N^{2/5}$,
since the plateau is mainly reached for $M\ge128$ and the corresponding
sizes of rings in monodisperse samples scale as $R\sim N^{2/5}$,
as discussed previously in ref \cite{Fischer}.

It will be very interesting to repeat the present analysis for bidisperse
blends of rings of clearly larger size such that the ratio of $N/M$
eventually becomes sufficient to change the plateau as function of
$M$ into an increasing function of $M$. If the analysis of ref.
\cite{Fischer} for larger $N$ is correct, there will be even more
interesting modifications to the behaviour at large $N$ and $M$.
First, if ring compression is sufficiently large, $N>N_{C}$, a non-knotting
contribution $N^{3}/R^{6}$ is balancing non-concatenation. For such
$N$, a different scaling of the plateau level should be observed
in Figure \ref{fig:Size-of-non-concatenated}. Next, above a critical
$N^{*}$ essentially any pair of overlapping rings should be entangled.
Then, compression should become a function of $M$ for sufficiently
large $N,M>N^{*}$ while remaining constant (beyond $N>N_{OO}$ and
up to the corrections mentioned above) for $N<N^{*}$.

\section{Summary\label{sec:Discussion}}

Ring conformations in entangled and interpenetrating melts were compared
in the present work. To understand the compression of the entangled
rings, the minimal area bounded by the ring contour is analyzed. It
is found that the area distribution of rings is decaying exponentially
as function of $(r_{b}-r_{c})/N^{\alpha}$, whereby $r_{b}$ is the
distance of the center of an area segment to the nearest monomer of
the ring, $r_{c}$ is a geometrical cut-off related to bead diameter,
and $N$ is the degree of polymerization of the ring. The characteristic
exponent is $\alpha\approx0.588$ for swollen rings, $\alpha\approx1/2$
for unentangled interpenetrating rings, and $\alpha\approx1/3$ for
entangled rings. The $N$-dependence of the exponential decay leads
to a correction $\sim N^{\alpha}$ to the minimal area bounded by
a ring polymer. For interpenetrating rings, this leads to a number
of rings that have to be expelled in case of monodisperse entangled
melts $\sim\phi R^{2}$, as found previously in ref. \cite{Fischer}.
The results of section \ref{sec:The-minimal-surface} demonstrate
that only the inner fraction of the area that is subject to penetrations
is getting modified upon compression of ring polymers in melts or
swelling of rings in low molecular weight solvents.

The internal structure of entangled rings was analyed to test the
applicability of the models available in literature \cite{Fischer,Sakaue2012,cates1986,Sakaue2011}.
An extension of the model of Cates and Deutsch \cite{cates1986} leads
to an onset of compression in melts with melt degree of polymerization
$M\sim N^{1/2}$, which is not confirmed by the simulation data of
the present study. The model of Sakaue \cite{Sakaue2012,Sakaue2011}
assumes that the topological length scale becomes a decreasing function
for increasing $M$. The model of Lang et al. \cite{Fischer} requires
a constant entanglement length and uses an exponential cut-off for
the entanglement effects towards small $M$. After this onset, the
effect of entanglements is assumed to be constant. The former two
models should manifest themselfes in a change of the internal structure
of the rings that is a function of melt molecular weight. The simulation
data of the present study agree only with the model of ref \cite{Fischer}:
for sufficiently large $M$ there is a broad range of melt molecular
weights with constant peak positions in the normalized mean square
internal distances $R_{e}(s)/s$ or in the bond-bond correlation function.
Both peaks arise only for entangled samples and are absent in interpenetrating
samples. Thus, we have to conclude that these peaks must be related
to entanglements between the rings. In consequence, the length scale
of topological interactions (entanglement length) in a melt of entangled
rings must be constant in contrast to the recent proposal by Sakaue
\cite{Sakaue2012}. This observation is corroborated by a comparison
of ring sizes in entangled and interpenetrating melts, which shows
that the ring size remains constant as soon as the concatenation probability
converges to one.

\section{Acknowledgement}

The author is indebted to J.-U. Sommer, J. Fischer, T. Kreer, and
M. Rubinstein for stimulating discussions on this subject and the
author thanks the DFG for funding grant LA2375/2-1 and the ZIH Dresden
for a generous grant of computing time under the project BiBPoDiA.

\bibliographystyle{JAmChemSoc_all}
\bibliography{literatur}

\section*{\newpage{}}
\section*{Table of Contents Graphic}
\emph{Minimal surface bounded by polymer rings: normalized distribution of area segments as function of the distance $r_b$ to the boundary}

Michael Lang

\begin{figure}[htbp]
\includegraphics[angle=0,width=\columnwidth]{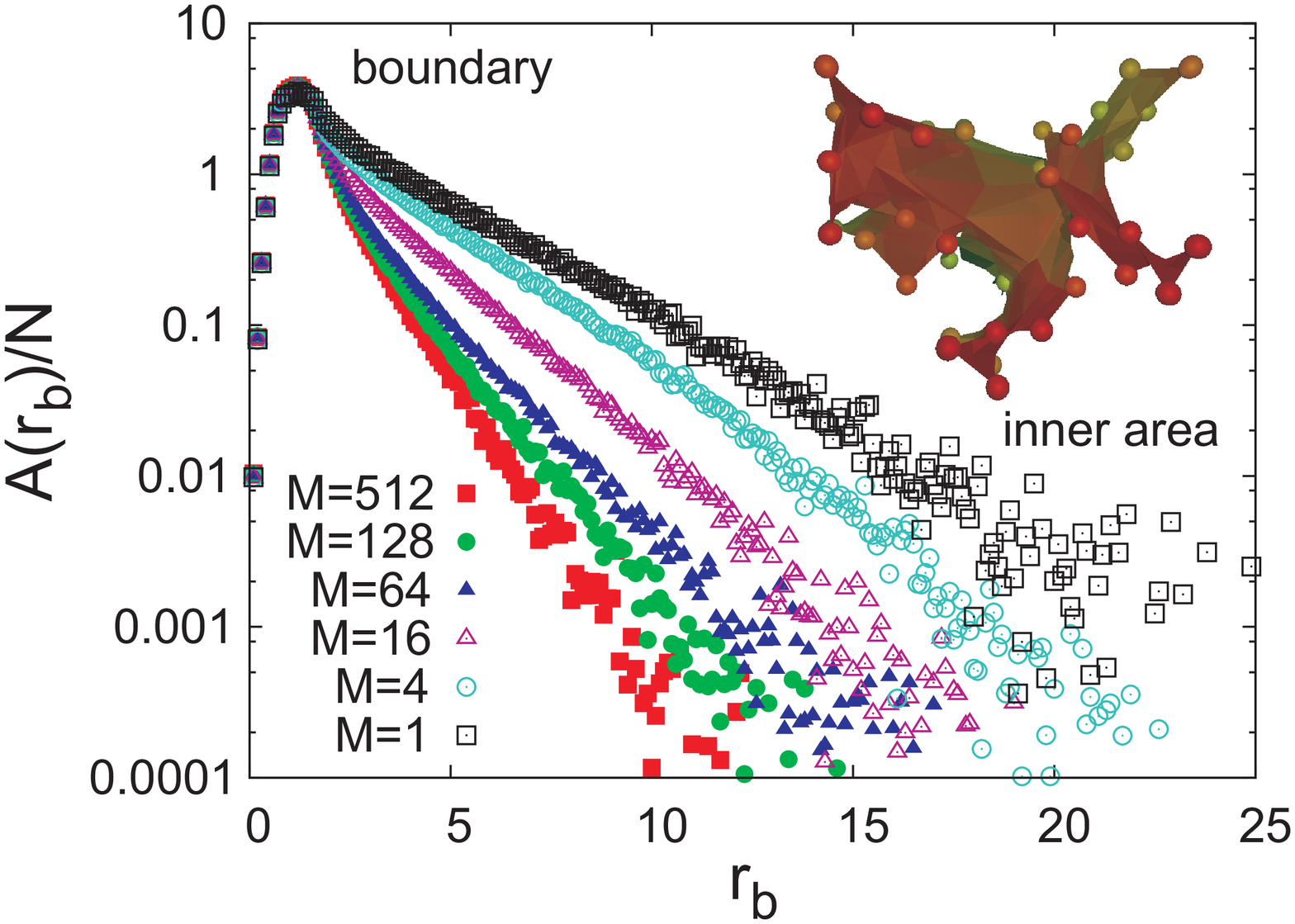}
\end{figure}
\end{document}